\def \SAIT #1 #2 {{\em Mem.\ Soc.\ Astron.\ It.\/} {\bf #1}, #2}
\def \MESS #1 #2 {{\em The Messenger\/} {\bf #1}, #2}
\def \ASTRNACH #1 #2 {{\em Astron. Nach.\/} {\bf #1}, #2}
\def \AAP #1 #2 {{\em Astron. Astrophys.\/} {\bf #1}, #2}
\def \AAL #1 #2 {{\em Astron. Astrophys. Lett.\/} {\bf #1}, L#2}
\def \AAR #1 #2 {{\em Astron. Astrophys. Rev.\/} {\bf #1}, #2}
\def \AAS #1 #2 {{\em Astron. Astrophys. Suppl. Ser.\/} {\bf #1}, #2}
\def \AJ #1 #2 {{\em Astron. J.\/} {\bf #1}, #2}
\def \ANNREV #1 #2 {{\em Ann. Rev. Astron. Astrophys.\/} {\bf #1}, #2}
\def \APJ #1 #2 {{\em Astrophys. J.\/} {\bf #1}, #2}
\def \APJL #1 #2 {{\em Astrophys. J. Lett.\/} {\bf #1}, L#2}
\def \APJS #1 #2 {{\em Astrophys. J. Suppl.\/} {\bf #1}, #2}
\def \APSS #1 #2 {{\em Astrophys. Space Sci.\/} {\bf #1}, #2}
\def \ASR #1 #2 {{\em Adv. Space Res.\/} {\bf #1}, #2}
\def \BAIC #1 #2 {{\em Bull. Astron. Inst. Czechosl.\/} {\bf #1}, #2}
\def \JSQRT #1 #2 {{\em J. Quant. Spectrosc. Radiat. Transfer\/} {\bf #1}, #2}
\def \MN #1 #2 {{\em Mon. Not. R. Astr. Soc.\/} {\bf #1}, #2}
\def \MEM #1 #2 {{\em Mem. R. Astr. Soc.\/} {\bf #1}, #2}
\def \PLR #1 #2 {{\em Phys. Lett. Rev.\/} {\bf #1}, #2}
\def \PASJ #1 #2 {{\em Publ. Astron. Soc. Japan\/} {\bf #1}, #2}
\def \PASP #1 #2 {{\em Publ. Astr. Soc. Pacific\/} {\bf #1}, #2}
\def \NAT #1 #2 {{\em Nature\/} {\bf #1}, #2}

\documentstyle{memsait}
\input epsf.sty
\begin{opening}
\title{RADIATIVE FEEDBACK AND THE PHOTOEVAPORATION OF INTERGALACTIC CLOUDS}
\author{PAUL R. SHAPIRO$^{1,2}$, ALEJANDRO C. RAGA$^2$, GARRELT MELLEMA$^2$}
\institute{$^1$Department of Astronomy, University of Texas,
               Austin, TX 78712, USA\\
           $^2$Instituto de Astronom\'\i a-UNAM, Apdo Postal 70-264,
               04510 M\'exico D. F., M\'exico\\
           $^3$Stockholm Observatory, S-133 36 SaltsJ\"obaden, Sweden
}
\date{} % DO NOT INSERT ANY DATE HERE !!!
\end{opening}

\begin{document}

%\oddpagefooter{\sf Mem. S.A.It., Vol. ??, ??}{}{\thepage}
%\evenpagefooter{\thepage}{}{\sf Mem. S.A.It., Vol. ??, ??}
\oddpagefooter{}{}{} % LEAVE AS IT IS !
\evenpagefooter{}{}{} % LEAVE AS IT IS !
\ 
\bigskip

\begin{abstract}
The first sources of ionizing radiation to condense out of the dark and 
neutral IGM sent ionization fronts sweeping 
outward through their surroundings, overtaking other primordial gas-clouds and photoevaporating them. Results
are presented of the first gas dynamical simulations of this process,
including radiative transfer, along with some observational diagnostics.
\end{abstract}

\section{Ionization Fronts in the IGM}
The r\^ole of hydrogen molecules as cooling agents in primordial gas clouds,
necessary to form the first stars out of a gas of H and He with no heavier
elements, is intimately related to the fate of those clouds in the presence
of ionizing and dissociating radiation. That such radiation had an important
effect on these primordial clouds is certain. The neutral, opaque IGM
out of which the first bound objects condensed was dramatically reheated
and reionized at some time between a redshift $z\approx50$ and $z\approx5$
by the radiation released by some of these objects
(cf. [1] and references therein). When the first sources
turned on, they ionized their surroundings by propagating weak, R-type
ionization fronts which moved outward supersonically with respect to both
the neutral gas ahead of and the ionized gas behind the front, racing ahead
of the hydrodynamical response of the IGM [2,3]. 
The effect of density inhomogeneity on the rate of I-front propagation
was previously described by a mean ``clumping factor'' $c_l>1$, which
slowed the I-fronts by increasing the average recombination rate per H atom
inside clumps. This suffices to describe the rate of I-front propagation
as long as the
clumps are either not self-shielding or, if so, only
absorb a fraction of the ionizing photons emitted by the central source. 
What is the dynamical effect of the I-front on the density inhomogeneity it
encounters, however? 

The answer depends on the size and density of
the clumps overtaken by the I-front.
The fate of linear density fluctuations depends upon their Jeans number, 
$L_J\equiv\lambda/\lambda_J$, the wavelength in units of the baryon Jeans 
length in the IGM at temperatures of order $10^4\rm K$. Fluctuations with
$L_J<1$ find their growth halted and reversed (cf. [4]).
For nonlinear density fluctuations, however, the answer is
more complicated, depending upon at least three dimensionless parameters,
their internal Jeans number, $L_J\equiv R_c/\lambda_J$, the ratio of
the cloud radius $R_c$ to the Jeans length $\lambda_J$ 
inside the cloud at about 
$10^4{\rm K}$, their ``Str\"omgren number'' $L_s\equiv R_c/\ell_s$, the
ratio of the cloud radius $R_c$ to the Str\"omgren length $\ell_s$ inside
the cloud (the length of a column of gas within which the unshielded arrival
rate of ionizing photons just balances the total recombination rate), and
their optical depth to H ionizing photons at 13.6 eV, $\tau_{\rm H}$,
before ionization. If $\tau_{\rm H}<1$, the I-front sweeps across the cloud, 
leaving an ionized gas at higher pressure than its surroundings, and exits
before any mass motion occurs in response, causing the cloud
to blow apart. If $\tau_{\rm H}>1$ and $L_s>1$, 
however, the cloud shields itself against ionizing photons,
trapping the I-front which enters the cloud, causing it to decelerate
inside the cloud to the sound speed of the ionized gas before it can exit
the other side, thereby transforming itself into a weak, D-type front
preceded by a shock. Typically, the side facing the source expels a supersonic
wind backwards towards the source, which shocks the IGM outside the cloud,
while the remaining neutral cloud material is accelerated away from the
source by the so-called ``rocket effect'' as the cloud photoevaporates
(cf. [5]).
As long as $L_J<1$ (the case for 
gas bound to dark halos with virial velocity less than
$\rm 10\,km\,s^{-1}$),
this photoevaporation proceeds unimpeded by 
gravity. For halos with higher virial velocity, however,
$L_J>1$, and gravity competes with pressure forces. For a
uniform gas of H density $n_{\rm H,c}$, located a distance $r_{\rm Mpc}$
(in Mpc) from a UV source emitting $N_{\rm ph,56}$ ionizing photons
(in units of $\rm10^{56}s^{-1}$), the Str\"omgren length is only
$\ell_s\cong(50\,{\rm pc})
(N_{\rm ph,56}/r_{\rm Mpc}^2)(n_{\rm H,c}/0.1\,\rm cm^{-3})^{-2}$.
We focus in what follows on the self-shielded case which traps the I-front.
Some of these results were summarized previously
by us in [6].

\section{The Photoevaporation of an Intergalactic Cloud Overtaken by a
Cosmological Ionization Front}

As a first study of these important effects, we have simulated the
photoevaporation of a uniform, spherical, neutral, intergalactic cloud of 
gas mass $1.5\times10^6M_\odot$, radius $R_c=0.5\,\rm kpc$, 
density $n_{\rm H,c}=0.1\,{\rm cm^{-3}}$ and $T=100\,\rm K$,
in which self-gravity is unimportant, located 
$1\,\rm Mpc$ from a quasar with emission spectrum $F_\nu\propto\nu^{-1.8}$
($\nu>\nu_{\rm H}$) and $N_{\rm ph}=10^{56}{\rm s}^{-1}$, initially
in pressure balance with
an ambient IGM of density $0.001\,\rm cm^{-3}$ which
at time $t=0$ has
just been photoionized by the passage of
the intergalactic R-type I-front generated when the quasar turned on.
[A standard top-hat perturbation which collapses 
and virializes at $z_{\rm coll}=9$,
for example, with total mass $\cong10^7M_\odot$, has circular velocity
$v_c\cong7\,\rm km\,s^{-1}$, $R_c\cong560\rm\,pc$, and
$n_{\rm H,c}=0.1\,\rm cm^{-3}$, if $\Omega_bh^2=0.03$ and $h=0.5$.]
Apart from H and He, the cloud also contains heavy elements at $10^{-3}$
times the solar abundance. Our simulations in 2D, axisymmetry use an
Eulerian hydro code (called CORAL), with Adaptive Mesh Refinement and
a Riemann solver based on the Van~Leer flux-splitting algorithm, which solves
nonequilibrium ionization rate equations (for H, He, C, N, O, Ne, and~S)
and includes an explicit treatment of radiative transfer
by taking into account the bound-free opacity of H and He [7,8,9].
Our grid size in cylindrical coordinates
$(r,x)$ was $128\times512$~cells (fully refined).

\begin{figure}
\epsfxsize=12cm % fix the x-dimension and scales y-dim. to x-dim.
% Feel free to do the choice you prefer but do not exceed the x-dimension
% of the text lines
\hspace{0.5cm}\epsfbox{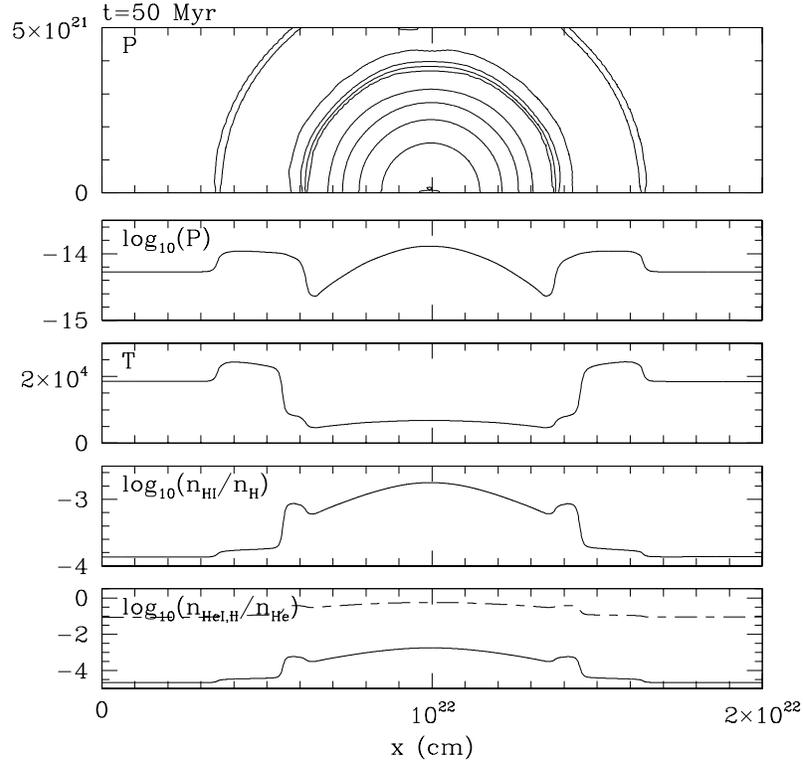} %for centering: act on hspace argument 
\vskip-1cm
\caption[h]{ZERO OPTICAL DEPTH APPROXIMATION.
One time-slice 50 Myr after turn-on of quasar located 1 Mpc away
from cloud to the left of computational box
along the $x$-axis. From top to bottom:
(a) isocontours of pressure, logarithmically spaced, in $(r,x)-$plane
of cylindrical coordinates; (b) pressure along the $r=0$ symmetry axis;
(c) temperature; (d) H~I fraction; (e) He~I (solid) and
He~II (dashed) fractions.}
\end{figure}

\begin{figure}
\epsfxsize=12cm % fix the x-dimension and scales y-dim. to x-dim.
% Feel free to do the choice you prefer but do not exceed the x-dimension
% of the text lines
\hspace{0.5cm}\epsfbox{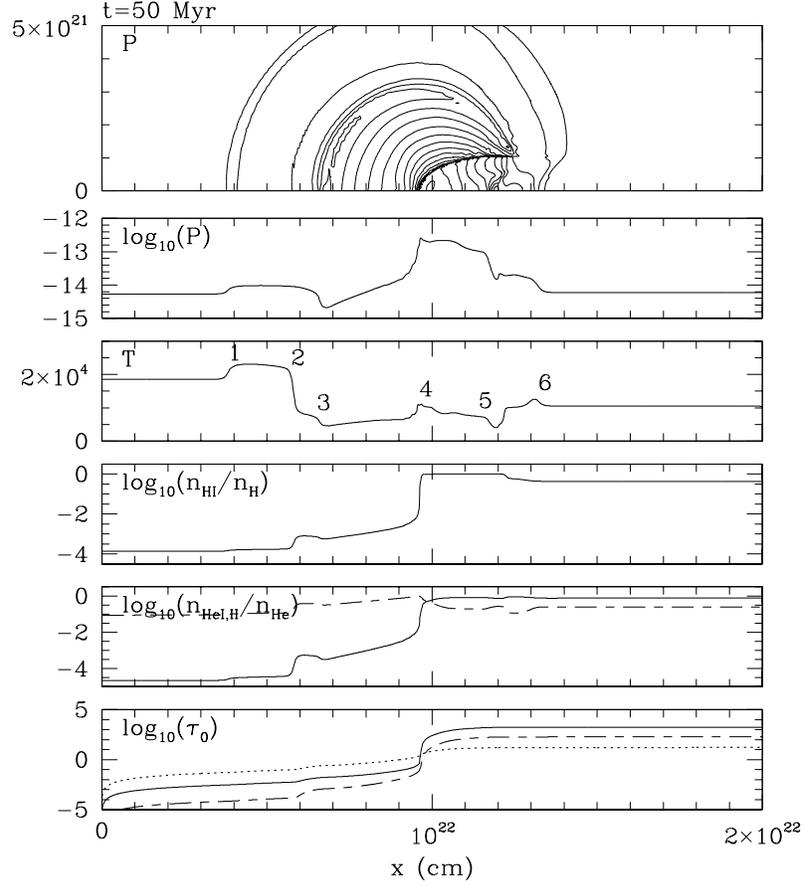} %for centering: act on hspace argument
\caption[h]{Same as Figure 1,
except: OPTICAL DEPTH INCLUDED. 
Bottom panel is bound-free optical depth along $r=0$ axis
at the threshold ionization energies for H~I (solid), He~I (dashed), He~II (dotted).}
\end{figure}

\begin{figure}
\epsfysize=10cm % fix the y-dimension and scales x-dim. to y-dim.
%\epsfxsize=8cm % fix the x-dimension and scales y-dim. to x-dim.
% Feel free to do the choice you prefer but do not exceed the x-dimension
% of the text lines
\hspace{1.5cm}\epsfbox{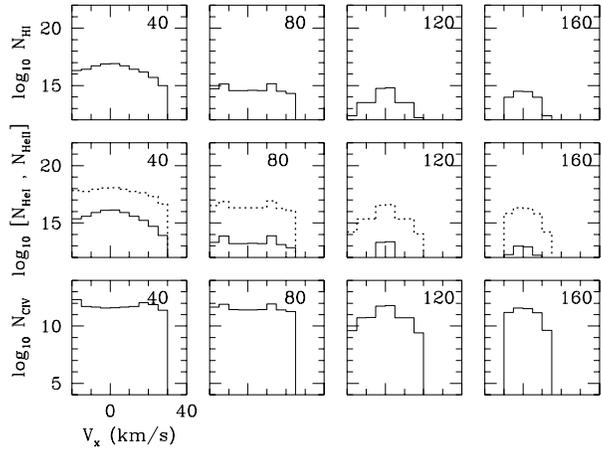} %for centering: act on hspace argument 
\vskip-2cm
\caption[h]{Cloud column densities ($\rm cm^{-2}$) along symmetry axis at
different velocities: ZERO OPTICAL DEPTH APPROXIMATION.
From top to bottom: (top) H~I; (middle) He~I (solid) and
He~II (dotted); (bottom) C~IV. Each box labelled with time (in Myrs) 
since QSO turn-on.}
\end{figure}

\begin{figure}
\epsfysize=10cm % fix the y-dimension and scales x-dim. to y-dim.
%\epsfxsize=8cm % fix the x-dimension and scales y-dim. to x-dim.
% Feel free to do the choice you prefer but do not exceed the x-dimension
% of the text lines
\hspace{1.5cm}\epsfbox{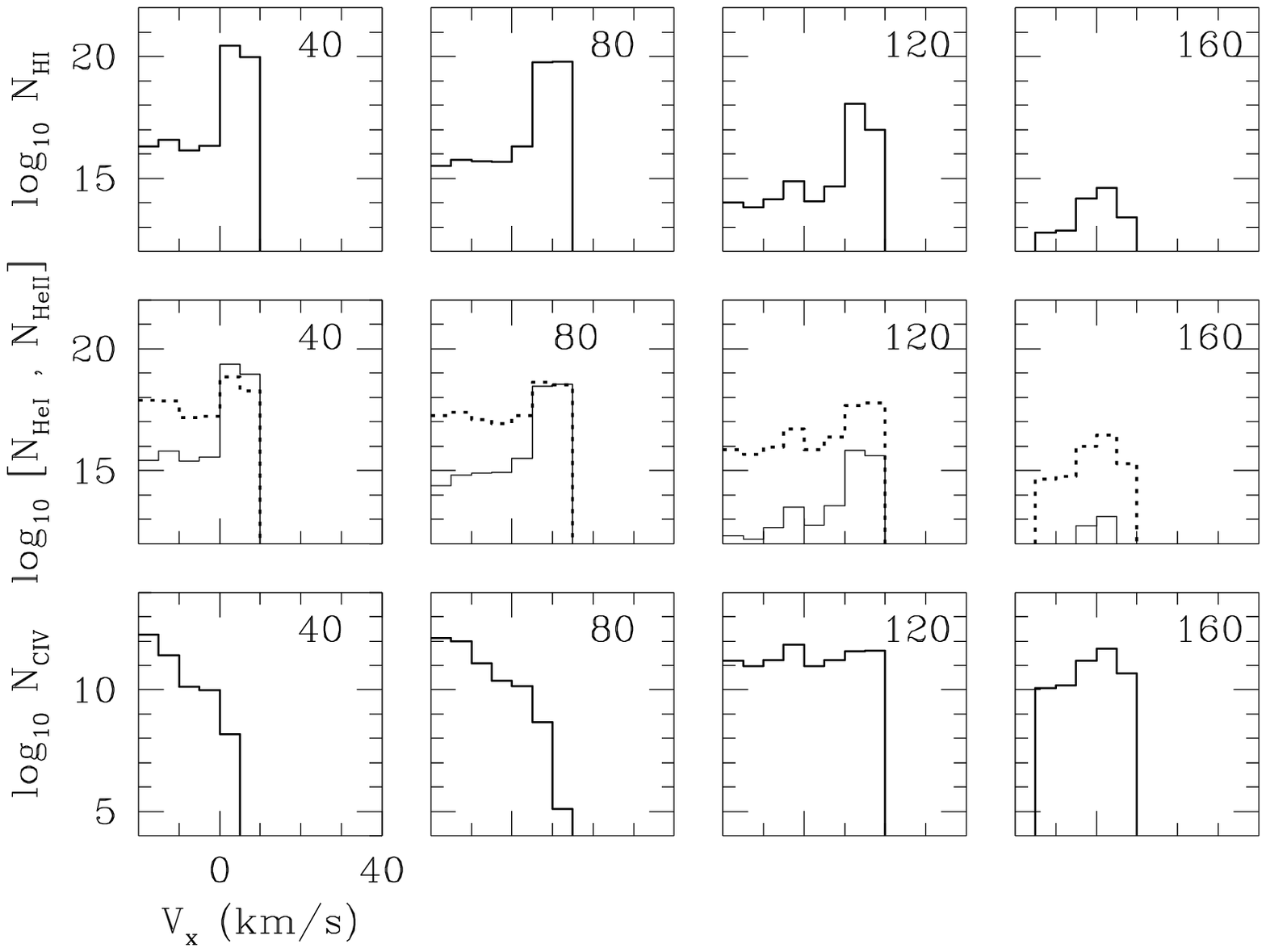} %for centering: act on hspace argument 
\vskip-2cm
\caption[h]{Same as Figure 3, except OPTICAL DEPTH INCLUDED.}
\end{figure}

Most current simulations of cosmological gas dynamics which include
photoionization do so by approximating the radiation field as uniform and 
isotropic, ignoring the inhomogeneity of the radiation field and setting
optical depth to zero. To illustrate the importance of a more realistic
approach to radiative transfer, we present results for two cases in what
follows: (1) zero optical depth and (2) optical depth properly included.
Figure~1 shows the structure of the cloud $50\,\rm Myr$ after
it was overtaken by the quasar's I-front as it sweeps past the cloud
in the IGM when optical depth is neglected. The cloud and IGM are both 
instantaneously photoionized everywhere in this case, causing the
overpressured cloud to expand isotropically into the surrounding IGM,
acting as a 
spherical piston which sweeps up the IGM and drives a shock into it.
The cloud matter eventually expands as a shell and evacuates a spherical hole.
The column densities of H~I, He~I and II, and C~IV for cloud gas of
different velocities as seen along the symmetry axis at different times
are shown in Figure 3. This cloud would initially resemble a 
velocity-broadened Lyman alpha forest quasar absorber (``LF'')
10's of $\rm km\,s^{-1}$ wide, with $N_{\rm H\,I}>10^{16}\rm cm^{-2}$, which
evolves toward a narrower LF absorber with $N_{\rm H,I}<10^{15}\rm cm^{-2}$,
with $N_{\rm He\,II}/N_{\rm H\,I}\leq10^2$ and 
$N_{\rm C\,IV}\sim10^{12}\rm cm^{-2}$ throughout. As the spatial variations for
the relative abundances of selected metal ions at 50 Myr plotted in 
Figure 5 show, only highly ionized metals are present in this case.
By contrast, we show results for all these same quantities 
from the simulation which takes proper account
of optical depth, instead, in Figures 2, 4, and 6. 
Since $\ell_S\ll R_c$ initially, the cloud traps the I-front,
as described above, and drives a supersonic wind from the surface facing the
quasar. It takes more than $100\,\rm Myr$ to evaporate the cloud,
accelerating it to 10's of $\rm km\,s^{-1}$ in the process.
Key features
of the flow are indicated by the numbers which label them on the
temperature plot in Figure 2: 1 = IGM shock; 2 = contact discontinuity
between shocked cloud wind and swept up IGM; 3 = wind shock; 
between 3 and 4 = supersonic wind; 4 = I-front; 5 = shock preceding I-front;
6 = shock that leads the motion of remaining neutral cloud gas
into the shadow region. At early times, the cloud gas resembles a weak
Damped Lyman Alpha (``DLA'') absorber with small velocity width 
($\sim10\rm\,km\,s^{-1}$) and $N_{\rm H\,I}\sim10^{20}\rm cm^{-2}$,
with velocity-broadened LF-like wings ($\hbox{width}\,\sim20\,\rm km\,s^{-1}$)
with $N_{\rm H\,I}\sim10^{16}\rm cm^{-2}$ on the side moving toward
the quasar, with a C~IV feature with
$N_{\rm C\,IV}\sim10^{12}\rm cm^{-2}$ displaced in
this same asymmetric way from the velocity of peak H~I column
density. After 160 Myr,
however, only a narrow H~I feature with LF-like column density
$N_{\rm H\,I}\sim10^{14}\rm cm^{-2}$ remains, with 
$N_{\rm He\,II}/N_{\rm H\,I}\sim10^2$ and
$N_{\rm C\,IV}/N_{\rm H\,I}\sim\rm[C]/[C]_\odot$. Unlike the zero-optical-depth
approximation, Figure 6 shows the presence at 50 Myrs
of low as well as high
ionization stages of the metals.

% To insert postscript versions of figures use the following commands:
% YOU NEED THE eps.sty FILE AND THE POSTSCRIPT FILES OF THE FIGURES
% IN THE SAME DIRECTORY.

% To insert harcopy versions of figures use the following commands:
% FIGURES MUST BE MOUNTED IN THE APPROPRIATE SPACES BEFORE
% SUBMITTING THE CAMERA-READY HARDCOPY.

\begin{figure}
\epsfysize=10cm % fix the y-dimension and scales x-dim. to y-dim.
%\epsfxsize=8cm % fix the x-dimension and scales y-dim. to x-dim.
% Feel free to do the choice you prefer but do not exceed the x-dimension
% of the text lines
\hspace{12.0cm}\epsfbox{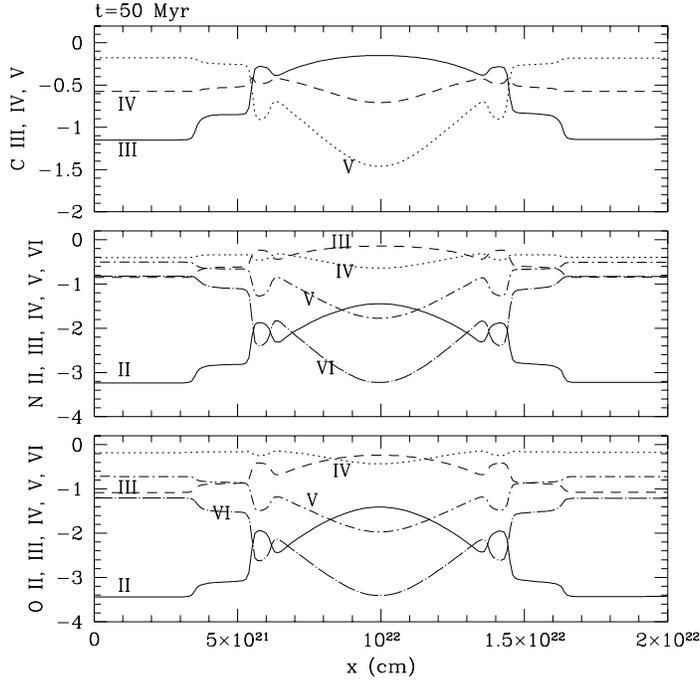} %for centering: act on hspace argument 
\caption[h]{Carbon, Nitrogen, and Oxygen Ionic Fractions Along Symmetry
Axis at $t = 50\rm\,Myr$: ZERO OPTICAL DEPTH APPROXIMATION.}
\end{figure}

\begin{figure}
\epsfysize=10cm % fix the y-dimension and scales x-dim. to y-dim.
%\epsfxsize=8cm % fix the x-dimension and scales y-dim. to x-dim.
% Feel free to do the choice you prefer but do not exceed the x-dimension
% of the text lines
\hspace{12.0cm}\epsfbox{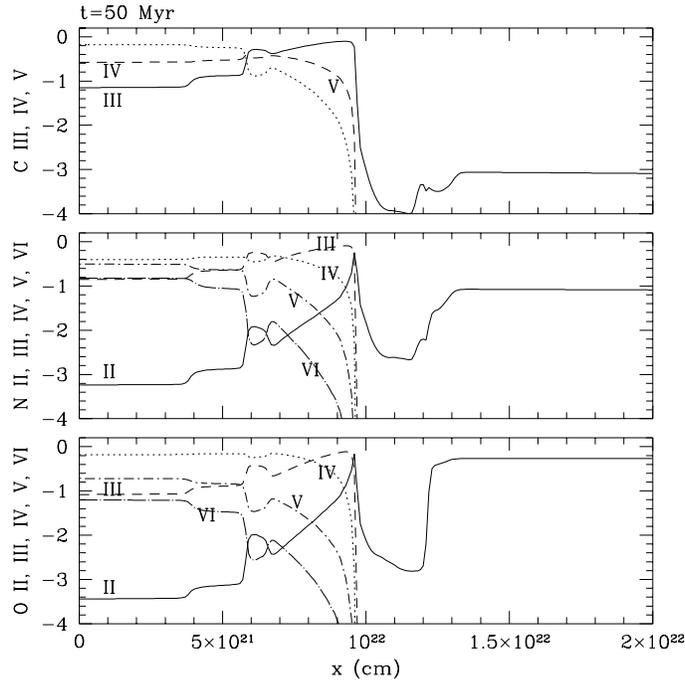} %for centering: act on hspace argument 
\caption[h]{Same as Figure 5, except OPTICAL DEPTH INCLUDED.}
\end{figure}

\acknowledgements
This work was supported by NASA Grant NAG5-2785 
and NSF Grant ASC-9504046, and was made possible by a UT Dean's
Fellowship and a National Chair of Excellence, UNAM, Mexico 
awarded by CONaCYT in 1997 for PRS.

% References. We avoided using the \bibitem commmand since we found it is
% somewhat platform-dependent. We also avoided using the \cite{keyword}
% command since we found it cumbersome. However, if you are an expert 
% LateX user you may use the various LateX tools for the references 
% provided they give the same printout formats of the examples given here.

\end{document}